\documentclass[fleqn,10pt]{wlscirep}

\usepackage{MnSymbol}
\usepackage{graphicx}

\title{Light-induced atomic desorption in a compact system for ultracold atoms}

\author[1,+]{Lara~Torralbo-Campo}
\author[1]{Graham~D.~Bruce}
\author[1]{Giuseppe~Smirne}
\author[1,*]{Donatella~Cassettari}
\affil[1]{Scottish Universities Physics Alliance, School of Physics and Astronomy, University of St~Andrews, North Haugh, St~Andrews, Fife KY16~9SS, United Kingdom}

\affil[*]{dc43@st-andrews.ac.uk}

\affil[+]{Current affiliation: Van der Waals-Zeeman Institute / Institute of Physics, University of Amsterdam, Science Park 904, 1098 XH Amsterdam}

\begin{abstract}
In recent years, light-induced atomic desorption (LIAD) of alkali atoms from the inner surface of a vacuum chamber has been employed in cold atom experiments for the purpose of modulating the alkali background vapour. This is beneficial because larger trapped atom samples can be loaded from vapour at higher pressure, after which the pressure is reduced to increase the lifetime of the sample. We present an analysis, based on the case of rubidium atoms adsorbed on pyrex, of various aspects of LIAD that are useful for this application. Firstly, we study the intensity dependence of LIAD by fitting the experimental data with a rate-equation model, from which we extract a correct prediction for the increase in trapped atom number. Following this, we quantify a figure of merit for the utility of LIAD in cold atom experiments and we show how it can be optimised for realistic experimental parameters.
\end{abstract}
\begin{document}

\flushbottom
\maketitle

\thispagestyle{empty}

\section*{Introduction}

Since the first realisation of Bose-Einstein condensation in ultracold atoms, a range of methods has been developed to produce quantum degenerate gases. Techniques are by now well established and many experiments have, as a starting point, the loading of laser-cooled atoms from background vapour in a magneto-optical trap (MOT), followed by evaporative cooling to quantum degeneracy in a conservative trap \cite{Ketterle99}. In order to cool and trap the required atom number, it is important to have a high partial background pressure during the MOT stage because a high percentage of atoms will be lost during evaporative cooling. However, evaporative cooling requires a lower background pressure than for the MOT loading, so that collisions with background atoms and molecules are minimised and the lifetime is sufficiently long. These two opposite constraints require a compromised solution. This is often a dual-chamber vacuum setup, where the process of MOT loading is spatially separated from the subsequent evaporation. More recently, with the development of atom chips \cite{Fortagh2007,Reichel} and all-optical evaporation \cite{Barrett2001,Thomas2001, Weiss2005, Clement2009, Jacob2011, Bruce11-2}, fast production of Bose-Einstein condensates has been demonstrated. This considerably relaxes the demand for long trap lifetimes and opens the possibility of designing vacuum systems based on single chambers, rather than dual chambers. This cuts down the size and complexity of the apparatus, which is beneficial for technological applications. 

Single chambers however reintroduce the original dilemma between atom number and lifetime. In this context, it is clearly advantageous to modulate the partial pressure of the atomic gas so that a large MOT is loaded, whilst keeping the partial pressure low during evaporation. Several techniques have been developed for this purpose, for instance pulsing alkali-metal dispensers \cite{Fortagh98,Rapol01,Moore05,Griffin05,McDowall12,Dugrain14}, or using weak non-resonant light to desorb atoms from the walls of the vacuum chamber\cite{Anderson01, tomassetti03, Arlt06, Telles10, Gerbier10, Thywissen05, Zhang09}. The latter, known as light-induced atomic desorption (LIAD)\cite{gozzini93}, can be seen as the atomic analogue of the photoelectric effect. The light source is pulsed for the length of time needed to load the MOT (typically between a few seconds up to tens of seconds), after which the light is turned off and the partial pressure drops back to a lower value. The lifetime recovered after the light pulse has been shown to be long enough to allow evaporation to quantum degeneracy\cite{Hansel01, Anderson04, Jacob2011}.

  \begin{figure}[hbt]
	\centering
		\includegraphics[width=0.45\textwidth]{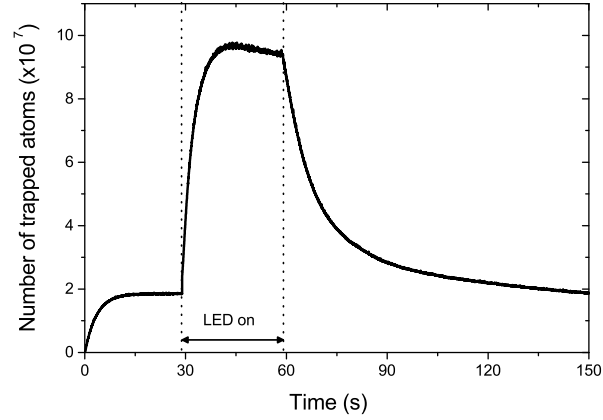}
	\caption{\textbf{The LIAD effect.} Atoms trapped in a Rb MOT during a violet light pulse (delimited by the dotted lines). LIAD increases the partial rubidium pressure, leading to a MOT of $\sim 10^{8}$ atoms.}
	\label{fig:N}
\end{figure}

Our investigation involves a MOT of $^{87}$Rb atoms created in a pyrex cell illuminated by light emitting diodes (LEDs) producing up to 9~mW/cm$^{-2}$ of violet light at a central wavelength of 405~nm (see Methods). The main motivation of our work is to develop a more quantitative understanding of the effectiveness of LIAD in cold atom experiments. As shown in Fig. \ref{fig:N}, we measure a significant increase (about a factor 5) in the number of trapped atoms as the violet light is pulsed on the cell. In our experiment we use the MOT also to monitor the partial rubidium pressure\cite{Anderson01} (see Methods) and in particular we measure the dependence of partial pressure on light intensity. This dependence has previously been investigated by other groups\cite{Arlt06, Zhang09} and a theoretical model appropriate for experimental data taken in the low-intensity regime has been proposed \cite{Zhang09}. Here we propose a more general rate-equation model that encompasses the regime of light intensities strong enough to cause a saturation of the rubidium pressure, which is more appropriate for our experimental data.
 
Another experimental method introduced in this paper is the $N_{\text{eq}}-\tau$ plot, where the trapped atom number at equilibrium $N_{\text{eq}}$ is plotted against the $1/e$ loading time $\tau$ of the trap (which coincides with its lifetime). We find that this approach is convenient first to characterise the system at unmodulated background, and subsequently to quantify a figure of merit of LIAD. The latter is defined as the increase of the product $N_{\text{eq}}\tau$ relative to the case of constant background pressure\cite{Anderson01}. Given that we can increase the MOT atom number temporarily during the light pulse and recover a low partial pressure after the pulse, $N_{\text{eq}}\tau$ can be maximised.

The paper is organised as follows: firstly, we introduce the method of the $N_{\text{eq}}-\tau$ plot. Secondly, the dependence of LIAD on light intensity is analysed in terms of the rate-equation model. Finally, we consider questions that are specifically relevant for the application of LIAD to cold atom experiments, namely the MOT increase with light intensity and the optimisation of the above-mentioned figure of merit.

\section*{Results}

\subsection*{MOT characterization with unmodulated Rb background pressure}\label{sec:appli1}

The loading of atoms in a MOT from background gas is the result of the balance between the atom capture rate and the rate of loss mechanisms. Hence the trapped atom number $N$ obeys the equation \cite{Monroe90,Steane92}: 

\begin{equation}
\label{eq:loading2}
\frac{dN}{dt}=R - \frac {N}{\tau}=\alpha P_{\text{Rb}} -(\beta P_{\text{Rb}}+\gamma)N. 
\end{equation}

The first term on the right-hand side describes the loading rate $R=\alpha P_{\text{Rb}}$ at which atoms are captured from background gas. Here $\alpha$ is proportional to the trapping cross section and $P_{\text{Rb}}$ is the partial Rb pressure. The second term, containing $1/\tau=\beta P_{\text{Rb}}+\gamma$, corresponds to the loss rate. $\gamma$ is proportional to the non-Rb background pressure, while both $R$ and $\beta P_{\text{Rb}}$ (where $\beta$ is a loss coefficient) are proportional to the partial Rb pressure. In equation (\ref{eq:loading2}) we neglect losses due to inelastic two-body collisions within the trap. These losses are proportional to the density of the trapped atoms, and because our MOT is large enough to be in the constant density regime (due to photon re-absorption), the corresponding loss rate is constant. Typical values for this loss rate for MOTs similar to ours\cite{Arpornthip13} are significantly smaller than our value of $\gamma$. 

\begin{figure}[t]
\centering
\includegraphics[width=0.45\textwidth]{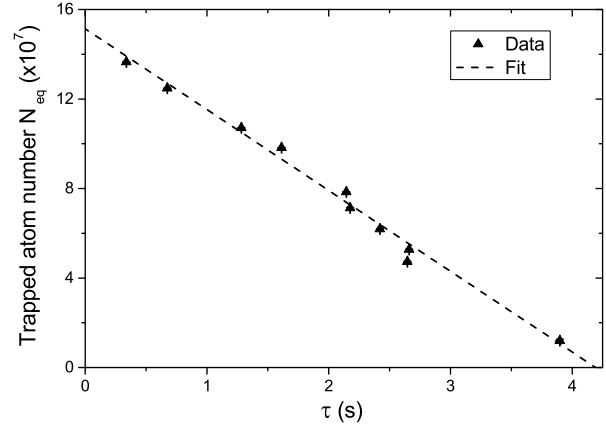}
\caption{\textbf{The $N_{\text{eq}}-\tau$ plot.} MOT atom number versus $1/e$ loading time $\tau$ in the case of unmodulated rubidium background. The fit with equation (\ref{eq:characterizationP}) provides values for the parameters $\alpha/\beta$ and $\gamma$. The error bars are uncertainties in the fit parameters of equation (\ref{eq:motEquation}).}
\label{fig:caract}
\end{figure}

From equation (\ref{eq:loading2}) we obtain the loading curve:

\begin{equation}
\label{eq:motEquation}
N(t)=N_{\text{eq}}[1 - e^{-t/\tau}],
\end{equation}

\noindent where $N_{\text{eq}}=\alpha P_{\text{Rb}}/(\beta P_{\text{Rb}} + \gamma)$ is the number of trapped atoms at equilibrium and $\tau$ is the $1/e$ loading time. The latter also indicates the lifetime of the atoms in the trap \cite{Anderson01,Anderson1994}. Given that both $\tau$ and $N_{\text{eq}}$ depend on the rubidium pressure $P_{\text{Rb}}$, it is convenient to eliminate $P_{\text{Rb}}$ and express $N_{\text{eq}}$ directly in terms of $\tau$ and the parameters $\alpha$, $\beta$ and $\gamma$:

\begin{equation}
\label{eq:characterizationP}
N_{\text{eq}}(\tau)=\frac{\alpha}{\beta}(1-\gamma\tau).
\end{equation}

In our experiment we measure the dependence of $N_{\text{eq}}$ on $\tau$, in the case of unmodulated background, by turning off the Rb dispensers after running them for several hours. We acquire several MOT loading curves as the Rb pressure gradually decays, and we fit each loading curve with equation (\ref{eq:motEquation}) to extract $N_{\text{eq}}$ and $\tau$. By fitting these data with equation (\ref{eq:characterizationP}) as shown in Fig. \ref{fig:caract}, we obtain $\alpha/\beta=(1.50\pm0.07)\times10^{8}$ and $\gamma=0.24\pm0.03$~s$^{-1}$. Physically the intercept with the vertical axis, $\alpha/\beta$, represents the largest MOT achievable in our chamber in the limit of partial Rb pressure much larger than the residual background from other gases. The intercept with the horizontal axis, $1/\gamma$, is the longest MOT lifetime achievable and is determined by the non-Rb background pressure. Hence this $N_{\text{eq}}-\tau$ plot is a useful method by which to characterise vacuum using cold atoms\cite{Moore14}.

LED pulses will temporarily increase the partial Rb pressure, but here we assume that they do not alter the parameters $\alpha/\beta$ and $\gamma$, which are therefore fixed and characteristic of our system. More specifically, for a MOT that is otherwise optimised, $\alpha/\beta$ is determined by the available optical power in the cooling laser beams, while $\gamma$ is given by the vacuum conditions. While it is conceivable that $\gamma$ may change during the LED pulses, i.e. the non-Rb background gases may also experience a LIAD effect, the measurements of LIAD-enhanced MOTs reported below are consistent with the assumption of constant $\gamma$. The parameters $\alpha/\beta$ and $\gamma$ will then play an important role in the figure of merit discussion.

\subsection*{LIAD dependence on LED intensity}

\begin{figure}[t]
	\centering
	\includegraphics[width=0.45\textwidth]{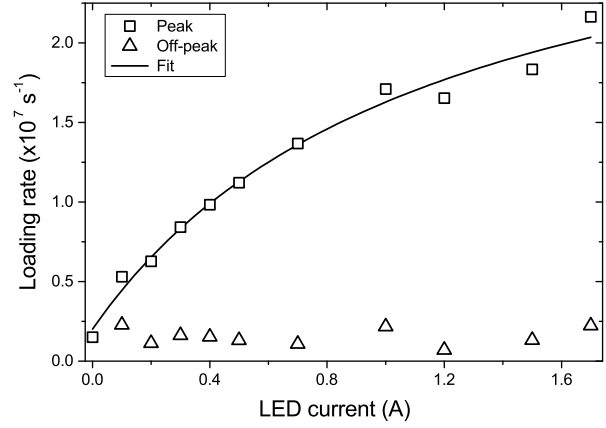}
	\caption{\textbf{LIAD-enhanced loading rate.} Loading rate measurements during the LED pulse (peak) and before the pulse (off-peak) as a function of LED current. The off-peak value is constant, which verifies that the variation is due to LIAD. The solid line is the fit of equation (\ref{eq:FitEquation}) to the data. 
	}
	\label{fig:pressLed}
	\end{figure}

To characterize LIAD in our system, we measure the MOT loading rate versus the LED current. The former is directly proportional to the partial rubidium pressure (see Methods) while the latter is directly proportional to the light intensity. For each value of LED current we measure the loading rate with LED on (off), which we refer to as peak (off-peak) loading rate. As shown in Fig. \ref{fig:pressLed}, the peak loading rate saturates at larger values of the LED current. To model this behaviour, which has also been observed in other experiments \cite{Arlt06}, we start from a rate equation for the number $N_{s}$ of rubidium atoms adsorbed on the surface of the glass cell. As discussed by Hatakeyama \emph{et al.} \cite{Hatakeyama06}, the typical partial Rb pressure in cold atoms experiments is so low that the surface density of adsorbed atoms is much less than a monolayer, i.e. adsorbed atoms are mainly isolated. In absence of LED light

\begin{equation}
\label{eq:ModelEquation0a}
\frac{dN_{s}}{dt}=-k_{d}N_{s}+k_{a}N_{v},
\end{equation}

\noindent where $N_{v}$ is the number of atoms in the volume of the cell, and $k_{d}$ and $k_{a}$ are the desorption and adsorption coefficients. The inverse of $k_{d}$ is referred to as sticking time $\tau_{s}$, i.e. the average time an atom sticks to the surface:  

\begin{equation}
\label{eq:ModelEquation0b}
\tau_{s}=\frac{1}{k_d}=\tau_{0}e^{E_{a}/k_{B}T},
\end{equation}

\noindent where $T$ is the surface temperature, $E_{a}$ is the adsorption energy and $\tau_{0}\sim 10^{-12}$s is the oscillation time of the bond \cite{deBoer}. The adsorption coefficient $k_{a}$ is proportional to the cell surface area, the flux of thermal atoms hitting the surface and the probability of an atom being adsorbed.

The steady-state solution of the rate equation (\ref{eq:ModelEquation0a}) is:

\begin{equation}
\label{eq:ModelEquation0d}
N_{v}=\frac{N_{t}}{1+\frac{k_a}{k_d}},
\end{equation}

\noindent where $N_{t}=N_{s}+N_{v}$ is the total number of atoms present in the cell, which we assume constant. This result is in agreement with the analysis presented by Stephens \emph{et al} \cite{Wieman94}. When the LED is turned on, a new equilibrium with increased $N_v$ is established faster than the time resolution of our pressure measurement (see Methods). By adding a desorption term $-kIN_s$ to the rate equation, where $I$ is the LED current and $k$ is a constant proportional to the LIAD cross section, we find a steady-state solution for this new equilibrium which we express directly in terms of the loading rate $R$ (proportional to $N_v$):

\begin{equation}
\label{eq:FitEquation}
R(I) \propto \frac{1+\frac{k}{k_{d}}I}{1+\frac{k_{a}}{k_{d}}+ \frac{k}{k_{d}}I}.
\end{equation}

\noindent We use this equation to fit the experimental data shown in Fig. \ref{fig:pressLed}, from which we obtain $k/k_{d}=18\pm5$ and $k_{a}/k_{d}=17\pm6$. 

Previous models\cite{Zhang09} have been used to fit the loading rate as a function of the light intensity in the linear range without including saturation. The main result of our analysis is that the observed saturation at large LED intensities emerges from the condition of constant $N_{t}$, and therefore can be explained by an effect of depletion of the surface rubidium. This is also consistent with our observation that over repeated LED pulses the rubidium atoms gradually leave the cell, an effect that we compensate for by running the Rb source at low current. 

Quantitatively the fit to the data provides information which, if combined with a microscopic characterisation of the surface, may lead to the determination of the adsorption energy and the LIAD cross-section. Rubidium atoms can be bound at the surface either to regular sites or to defect sites of silica. The latter provide a stronger bond and it has been suggested that the major contribution to LIAD is from atoms desorbed from defect sites, specifically from non-bridging oxygen (NBO) defects \cite{Hatakeyama06, Dominguez04}. If this is the case, the above rate-equation model could be applied to the adsorption and desorption of atoms specifically from the NBO centres. In particular the adsorption coefficient $k_a$ can be estimated if the surface density of NBO centres is known, and from this the coefficients $k_d$ and $k$ could be determined. However the occurrence of NBO centres is affected by the processing history of the glass surface \cite{Hatakeyama06} and is unknown for our uncharacterised cell. In the following, we use the Rb pressure increase measured in Fig. \ref{fig:pressLed} to quantify the corresponding increase in MOT atom number. Therefore from now on, our analysis is independent from the precise desorption mechanism.

\subsection*{Application to cold atom experiments: MOT dependence on LED intensity}\label{sec:appli}

\begin{figure}[ht]
	\centering
		\includegraphics[width=0.45\textwidth]{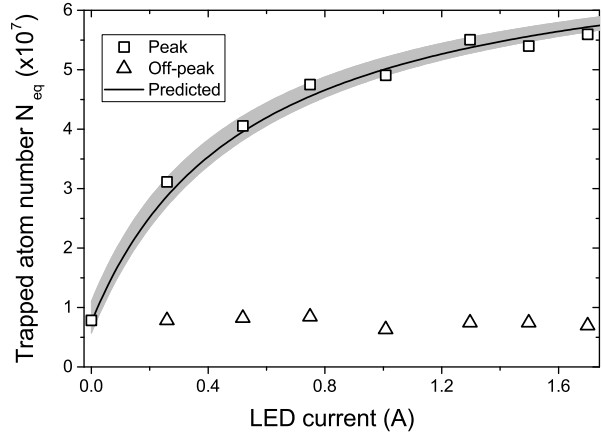}
	\caption{\textbf{LIAD-enhanced trapped atom number.} MOT atom number during the LED pulse (peak) and before the pulse (off-peak) versus LED current. The off-peak value is constant, which verifies that the variation in atom number is due to LIAD. The predicted atom number is calculated from equation (\ref{eq:characterizationP4}). 
The shaded area shows the extent of uncertainties in the parameters of equation (\ref{eq:characterizationP4}), as extracted from the fit in Fig. \ref{fig:pressLed}.}
	\label{fig:Ncurrent}
\end{figure}

As shown in Fig. \ref{fig:N}, during the LED pulse the number of atoms in the MOT increases from the off-peak value $N_{\text{eq}}^{\text{off}}=\alpha P_{\text{Rb}}/(\beta P_{\text{Rb}}+\gamma)$ to the peak value $N_{\text{eq}}^{\text{on}}=\alpha \eta P_{\text{Rb}}/(\beta\eta P_{\text{Rb}}+\gamma)$. Here $\eta(I)=R(I)/R(I=0)$ is the relative factor of increase in partial Rb pressure, which is proportional to the curve shown in Fig. \ref{fig:pressLed}. By eliminating $P_{\text{Rb}}$, we express $N_{\text{eq}}^{\text{on}}(I)$ in terms of $N_{\text{eq}}^{\text{off}}$ as:

\begin{equation}
\label{eq:characterizationP4}
N_{\text{eq}}^{\text{on}}(I)=\frac{\frac{\alpha}{\beta}\eta(I) N_{\text{eq}}^{\text{off}}}{\frac{\alpha}{\beta}+N_{\text{eq}}^{\text{off}}[\eta(I)-1]},
\end{equation}

\noindent and we use this equation to predict the peak atom number for different LED intensities, i.e. for different values of $\eta$. This is shown in Fig. \ref{fig:Ncurrent} alongside the experimental data. For the measurements, we pulse the LEDs at different currents for 30~s, which is more than sufficient to load full MOTs. The predicted curve is calculated using the measured off-peak $N_{\text{eq}}^{\text{off}}=7.5\times10^{6}$ atoms, $\alpha/\beta=1.50\times10^{8}$, and $\eta(I)$ from the fit to the data in Fig. \ref {fig:pressLed}. This shows good agreement with the experimental points. Therefore our model predicts how the MOT grows in presence of LIAD using an equation with no free parameters. 

Fig. \ref{fig:Ncurrent} shows LIAD-enhanced MOTs that are up to a factor 6 larger than the off-peak MOT.  We also find that moderate LED currents of about 1~A are sufficient to saturate the MOT atom number. A similar saturation effect was also observed by Klempt \emph{et al} \cite{Arlt06}. In our analysis this is due to the presence of an intensity-dependent term in the denominator of equation (\ref{eq:characterizationP4}). 

Finally we note that in order to derive equation (\ref{eq:characterizationP4}) from the expressions for $N_{\text{eq}}^{\text{off}}$ and $N_{\text{eq}}^{\text{on}}$, we rely on the previously-introduced assumption that $\gamma$ is the same in both expressions, i.e. that it is not affected by the LED pulse. Therefore the agreement between equation (\ref{eq:characterizationP4}) and the experimental data supports this assumption.

\subsection*{Optimisation of LIAD figure of merit}

Our chosen figure of merit of LIAD is the increase of the product $N_{\text{eq}}\tau$. Here we compare this product for the two cases of MOT loaded from constant Rb background and from LIAD-modulated Rb background.

In the modulated background case, we use the peak value for the MOT atom number and the off-peak value for the lifetime. This is justified because the off-peak lifetime is recovered soon after the LEDs are turned off. This recovery time varies among different experiments, but in most cases\cite{Gerbier10} the pressure drops to one tenth of its peak value over a period that ranges from 0.1 to 2~s. To proceed with evaporation, it is sufficient to keep the MOT on for this short period, during which most trapped atoms are retained. After this period, the pressure has recovered to essentially the same value as before the pulse and the atoms may be transferred to a conservative trap for subsequent evaporation.

\begin{figure}[ht]
	\centering
		\includegraphics[width=0.45\textwidth]{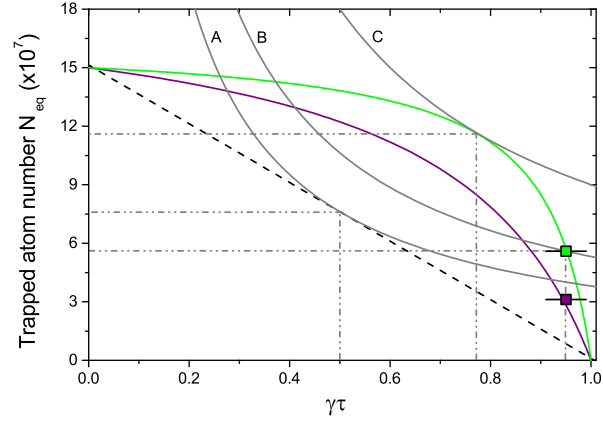}
		\caption{\textbf{$N_{\text{eq}}-\tau$ plot for LIAD-modulated Rb background.} MOT atom number versus $\gamma\tau$ for unmodulated Rb background (dashed straight line) and for LIAD-modulated Rb background (purple and green curves). The straight line is the fit to the data from Fig. \ref{fig:caract}, while the purple and green curves are given by equation (\ref{eq:characterizationP2}) at LED currents 0.25~A ($\eta=4.4$) and 1.7~A ($\eta=11.7$) respectively, for which experimental data are also taken (squares). The three hyperbolae define loci of constant $N_{\text{eq}}\gamma\tau$ (grey curves). The largest value of $N_{\text{eq}}\gamma\tau$ for unmodulated background is $\text{A}=3.81\times10^{7}$. The values $\text{B}=5.32\times10^{7}$ and $\text{C}=9.0\times10^{7}$ are for modulated background. B is the value attained experimentally, while C is the predicted upper limit for the LED intensity available in our experiment. The horizontal error bars on the experimental data come from the uncertainty in the determination of $\gamma$ from the fit in Fig. \ref{fig:caract}.}
	\label{fig:liadlife}
	\end{figure}

Hence we can extend the $N_{\text{eq}}-\tau$ plot shown in Fig. \ref{fig:caract} to include MOTs loaded with LIAD. Having both LIAD-enhanced MOTs and off-peak MOTs plotted versus the \itshape same \normalfont lifetime $\tau$ leads to a direct comparison between constant and modulated Rb background. This is shown in Fig. \ref{fig:liadlife}, where the horizontal axis is now the dimensionless product $\gamma\tau$.  

Starting from equation (\ref{eq:characterizationP4}), $N_{\text{eq}}^{\text{on}}$ is expressed in terms of $\gamma\tau$ by substituting equation (\ref{eq:characterizationP}) for $N_{\text{eq}}^{\text{off}}$. This leads to 

\begin{equation}
\label{eq:characterizationP2}
N_{\text{eq}}^{\text{on}}(\gamma\tau)=\frac{\alpha}{\beta}\left[1-\frac{\gamma\tau}{\eta(1-\gamma\tau)+\gamma\tau}\right],
\end{equation}

\noindent which is plotted in Fig. \ref{fig:liadlife} for two LED intensities, along with the unmodulated background line for comparison. The two experimental points on the plot show good agreement with the theoretical curves for the corresponding LED intensities. These data were taken at low off-peak Rb background, more specifically at an off-peak lifetime of $\tau=3.95$~s, which is close to the lifetime upper limit of $1/\gamma=4.17$~s. The resulting $\gamma\tau$ is 0.95.

Fig. \ref{fig:liadlife} also contains hyperbolae defined as $N_{\text{eq}}\gamma\tau=\text{const}$. If we compare the best possible $N_{\text{eq}}\gamma\tau$ value for unmodulated background (curve A) to the largest observed value for modulated background (curve B), we obtain a factor 1.4 increase. This is modest, however Fig. \ref{fig:liadlife} suggests that $N_{\text{eq}}\gamma\tau$ can be improved further by working at smaller $\gamma\tau$ values, i.e. at larger off-peak Rb background. This is the case of curve C, which corresponds to the best $N_{\text{eq}}\gamma\tau$ value for the highest LED intensity available in our experiment, and which leads to a factor 2.4 improvement relative to unmodulated background. 

\begin{figure}[ht]
	\centering
				\includegraphics[width=0.45\textwidth]{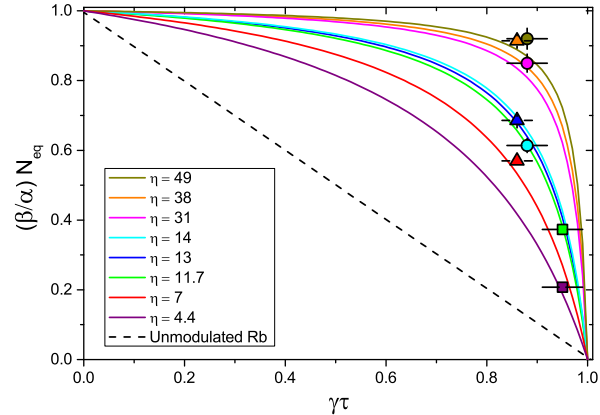}
	\caption{\textbf{$N_{\text{eq}}-\tau$ plot for data from different experiments.} $\circ$ denote the data of Telles \emph{et al.}\cite{Telles10}, $\medtriangleup$ denote the data of Mimoun \emph{et al.}  \cite{Gerbier10} and $\medsquare$ denote our data. Here the vertical axis is rescaled by $\beta/\alpha$ in order to have the data from different experiments on the same plot. The solid curves are calculated from equation (\ref{eq:characterizationP2}). The error bars come from the uncertainty in the determination of $\alpha/\beta$ and $\gamma$.}
	\label{fig:liadcomp}
\end{figure}

Given that the experimental data presented in this paper are taken in the limit of low off-peak Rb pressure, i.e. close to the upper limit for $\gamma\tau$, we apply our analysis to the LIAD data of previously reported experiments \cite{Telles10, Gerbier10}. For this purpose, firstly we extrapolate $\alpha/\beta$ and $\gamma$ from their data. From this we estimate that both these experiments work in a regime of smaller $\gamma\tau$. As predicted by equation (\ref{eq:characterizationP2}), larger improvements in $N_{\text{eq}}\gamma\tau$ are found experimentally (see Fig. \ref{fig:liadcomp}). The resulting factor of increase of $N_{\text{eq}}\gamma\tau$, relative to the unmodulated background, is 3.2 for Telles \emph{et al.} \cite{Telles10} and 3.1 for Mimoun \emph{et al.}\cite{Gerbier10} The fact that these experiments are carried out with different atomic species and in different vacuum conditions confirms the general applicability of our analysis to a wide range of experimental scenarios. 

\begin{figure}[t]
	\centering
		\includegraphics[width=0.45\textwidth]{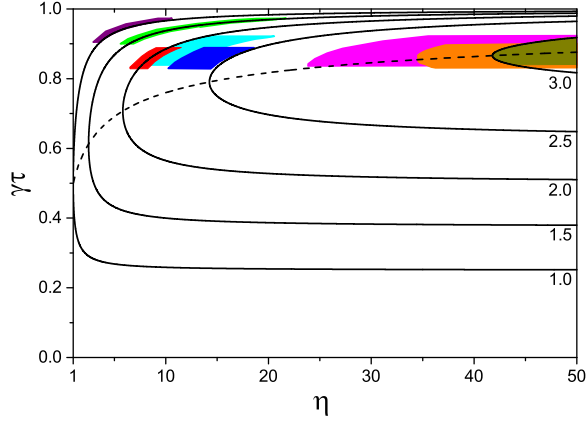}
	\caption{\textbf{Optimisation of the figure of merit.} Contours indicate the factor of increase of $N_{\text{eq}}\gamma\tau$ relative to unmodulated background. The dashed line indicates the optimal value $\left(\gamma\tau\right)_{\text{max}}$ as a function of $\eta$. The experimental data points from Fig. \ref{fig:liadcomp} are denoted by the shaded regions, which are the projections of the errors in $\gamma\tau$ and $N_{\text{eq}}$ onto the contour plot.}
	\label{fig:FOMoptim}
\end{figure}

Finally in the limit of very strong LED intensities we obtain a factor 4 increase in $N_{\text{eq}}\gamma\tau$, which corresponds to the ideal scenario where the MOT saturates at $\alpha/\beta$ during the LED pulse and a lifetime close to 1/$\gamma$ is recovered after the pulse\cite{Anderson01}. If one works at large values of $\eta$, it is therefore advantageous to choose $\gamma\tau$ close to 1. In general, equation (\ref{eq:characterizationP2}) can be used to calculate the optimal value of $\gamma\tau$ for a given $\eta$, i.e. the choice of $\gamma\tau$ that maximises the figure of merit. This is given by:

\begin{equation}
\left(\gamma\tau\right)_{\text{max}}=\frac{\eta-\sqrt{\eta}}{\eta-1},
\end{equation}

\noindent which is plotted as a dashed line in Fig. \ref{fig:FOMoptim}. The corresponding figure of merit, relative to the optimal $N_{\text{eq}}\gamma\tau=\alpha/ 4\beta$ for unmodulated background (hyperbola A in Fig. \ref{fig:liadlife}), is given by:

\begin{equation}
\frac{\left(N_{\text{eq}}\gamma\tau\right)_{\text{max}}} {\alpha/4\beta}=4 \frac{\eta}{\left(1+\sqrt{\eta}\right)^2}.
\end{equation}

We suggest that this analysis can be used as a guide for the best choice of $\gamma\tau$ in an experiment, given the available $\eta$. We note that a similar analysis can be performed for a different choice of the figure of merit of LIAD, for example one which emphasizes gains in atom number over lifetime. In general, however, working at $\gamma\tau$ between 0.8 and 0.9 is a good choice because with reasonable LED intensities the peak MOT can be close to $\alpha/\beta$ while long lifetimes are maintained.

\section*{Discussion}

We used a MOT to study light-induced desorption of rubidium atoms from pyrex. We developed a rate-equation model for the dependence on LED intensity, and we used this model to correctly predict the enhancement in the MOT atom number during the LED pulse. This led to a figure-of-merit analysis based on an $N_{\text{eq}}-\tau$ plot, where LIAD-enhanced MOTs are compared to constant-background MOTs. We found that for a factor 11.7 increase in rubidium pressure during the LED pulse (which was obtained at maximum LED intensity in our setup), the $N_{\text{eq}}\tau$ figure of merit increases by a factor 1.4 compared to the constant background case. At the same LED intensity, the factor of increase should be up to 2.4 when working at higher off-peak rubidium pressure. We also found that our model provides correct predictions for previously reported experimental results \cite{Telles10, Gerbier10}, for which the factor of increase in the figure of merit is closer to the theoretical limit of 4. Hence we suggest that this analysis, and the $N_{\text{eq}}-\tau$ plot more generally, may find broad applicability to cold atom experiments. In particular we expect it to be useful for the optimisation of experiments that use LIAD to improve $N_{\text{eq}}\tau$ in a single chamber setup. It may also be of interest for cooling and trapping radioactive atoms\cite{Aubin2003,Coppolaro2014}, where only a low vapour pressure or weak flux is available.

\section*{Methods}

\subsection*{Experimental setup}
\begin{figure}[ht]
\centering
\includegraphics[width=0.45\textwidth]{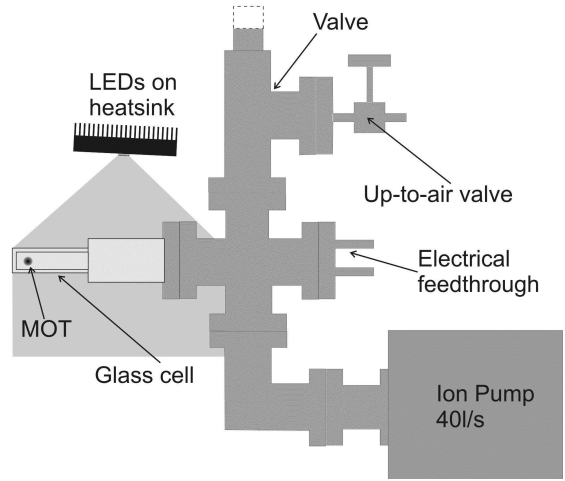}
\caption{\textbf{Single-chamber apparatus based on a glass cell.} The entire glass cell is illuminated by up to 9~mW/cm$^{2}$ from the 405~nm LEDs.}
\label{fig:glassCell3}
\end{figure}

The experimental setup is based on the compact vacuum system shown in Fig.~\ref{fig:glassCell3}. It consists of a CF40 four-way cross connected to a rectangular, uncoated pyrex glass cell of external dimensions 2.4$\times$2.4$\times$7.3~cm$^{3}$, a 40~l/s ion pump, an all-metal valve, and a 4-pin electrical feedthrough with two commercial Rb dispensers (SAES Getters RB/NF/7/25FT10+10). The dispensers are placed about 25~cm from the MOT trapping region and release rubidium atoms in the glass cell. The cooling and repumping laser light needed for the MOT is provided by two external-cavity diode lasers. The cooling light is tuned 14~MHz below the $5\text{S}_{1/2} (\text{F}=2) \rightarrow 5\text{P}_{3/2} (\text{F}'=3)$ transition and the repumper light is on resonance with the $5\text{S}_{1/2} (\text{F}=1) \rightarrow 5\text{P}_{3/2} (\text{F}'=2)$ transition. We have 40~mW and 5~mW of cooling and repumper power respectively. The laser beams are expanded and collimated to a beam waist of 7~mm and then split into six MOT beams. A calibrated photodetector-lens setup is used to collect the MOT fluorescence to measure the number of trapped atoms.

As light source for LIAD, we use an Enfis Uno Tag Array. This is a high-power surface-mounted device containing a 1~cm$^{2}$ array of 25 light emitting diodes. The LEDs emit violet light with a centre wavelength of 405~nm ($\text{FWHM}=16$~nm) and a maximum output power of 5~W, which can be controlled by the applied current. The device is mounted on a heatsink to dissipate the significant amount of heat produced, and is placed 13~cm away from the MOT region as shown in Fig.~\ref{fig:glassCell3}. The light from the LEDs is not collimated. 80$\%$ of the power falls within a 30$^\circ$ solid angle, providing an average intensity of 9~mW/cm$^{2}$ in the MOT region at  the maximum LED current of 1.7~A. The light intensity has a linear dependence on the LED current between 0 and 1.7~A. 

\subsection*{Monitoring the partial Rb pressure}

We use a method of measuring the partial Rb pressure in the MOT region similar to that used by Anderson and Kasevich \cite{Anderson01}. We take a sequence of repeated partial MOT loadings by applying a pulse train to the current through the MOT quadrupole coils, whereby the coils are on for 1~s and off for 3~s. As this pulse train is applied, we measure the number of trapped atoms after the 1~s of MOT loading. This is a direct measure of the MOT loading rate $R$, as can be seen from equation (\ref{eq:loading2}) with $N\rightarrow$0 and $dt=1$s. Given that the loading rate is proportional to the partial rubidium pressure, 
this method gives real-time monitoring of its evolution before (off-peak), during (peak) and after the LED pulse. We find that the rubidium pressure quickly reaches an equilibrium as the LEDs are turned on, stays approximately constant during the pulse, and decays after the pulse with a characteristic time constant. 

The loading rate and the rubidium pressure are linked by \cite{Monroe90, Steane92}

\begin{equation}
\label{eq:Pressure1}
P_{\text{Rb}}=k_{B}T \frac{v_{\text{th}}^3}{2 w_{0}^{2} v_{c}^{4}}R, 
\end{equation}

\noindent where $T$ is the room temperature, $v_{\text{th}}$ is the thermal velocity of the rubidium background, $w_{0}$ is the cooling laser beam waist and $v_{c}$ is the capture velocity of the MOT.  Using  $w_{0}=7$~mm and an estimated $v_{c}=20$~m/s, we obtain an off-peak value of $1.6\times10^{-10}$~mbar for the rubidium pressure. For the optimal case of 9~mW/cm$^{2}$ violet illumination, the rubidium pressure reaches $\sim10^{-9}$~mbar. After the violet pulse the partial Rb pressure goes back to the initial pressure of $1.6\times10^{-10}$~mbar.


\section*{Acknowledgements}

This work was supported by the UK EPSRC grant GR/T08272/01 and  the Leverhulme Trust Research Project Grant RPG-2013-074. G.S. acknowledges support from a SUPA Advanced Fellowship.

\end{document}